\documentclass[twocolumn, prl, superscriptaddress]{revtex4-1}
\usepackage{graphicx,amsmath}

\newcommand{\ua}{\uparrow}
\newcommand{\da}{\downarrow}

\begin{document}

\title{Effects of spin-orbit coupling and spatial symmetries\\ on the Josephson current in SNS junctions}
\author{Asbj\o rn Rasmussen}
\affiliation{Center for Quantum Devices, Niels Bohr Institute, University of Copenhagen, 2100 Copenhagen, Denmark}
\author{Jeroen Danon}
\affiliation{Center for Quantum Devices, Niels Bohr Institute, University of Copenhagen, 2100 Copenhagen, Denmark}
\affiliation{Niels Bohr International Academy, Niels Bohr Institute, University of Copenhagen, 2100 Copenhagen, Denmark}
\author{Henri Suominen}
\affiliation{Center for Quantum Devices, Niels Bohr Institute, University of Copenhagen, 2100 Copenhagen, Denmark}
\author{Fabrizio Nichele}
\affiliation{Center for Quantum Devices, Niels Bohr Institute, University of Copenhagen, 2100 Copenhagen, Denmark}
\author{Morten Kjaergaard}
\affiliation{Center for Quantum Devices, Niels Bohr Institute, University of Copenhagen, 2100 Copenhagen, Denmark}
\author{Karsten Flensberg}
\affiliation{Center for Quantum Devices, Niels Bohr Institute, University of Copenhagen, 2100 Copenhagen, Denmark}

\begin{abstract}
We present an analysis of the symmetries of the interference pattern of critical currents through a two-dimensional superconductor-semiconductor-superconductor junction, taking into account Rashba and Dresselhaus spin-orbit interaction, an arbitrarily oriented magnetic field, disorder, and structural asymmetries.
We relate the symmetries of the pattern to the absence or presence of symmetries in the Hamiltonian, which provides a qualitative connection between easily measurable quantities and the spin-orbit coupling and other symmetries of the junction. We support our analysis with numerical calculations of the Josephson current based on a perturbative expansion up to eighth order in tunnel coupling between the normal region and the superconductors.
\end{abstract}

\maketitle

%\section{Introduction}
Semiconductors with strong spin-orbit interaction (SOI) attracted a lot of attention in recent years. The prospect of manipulating electron spin efficiently with electric fields instead of magnetic fields makes SOI an attractive ingredient for spintronic applications~\cite{dattadas,spinrev}, as well as spin-based quantum computing~\cite{flindt:prl,stevan:nature}.  Furthermore, several concrete proposals were put forward on how to create topological states of matter in hybrid structures relying on semiconductors with strong SOI:  One- or two-dimensional semiconductors proximitized by an $s$-wave superconductor can behave as a $p$-wave topological superconductor~\cite{PhysRevLett.105.177002,PhysRevLett.105.077001,2dsau,2dalicea}. Two-dimensional semiconductor heterostructures can acquire an ``inverted band structure'' and enter a (topological) quantum spin Hall state~\cite{bernevig_2006,PhysRevLett.100.236601}. The notion that such topological systems can host non-Abelian quasiparticles and the prospect of using these particles for topologically protected quantum computing~\cite{RevModPhys.80.1083} sparked an intense activity of research and fueled the interest in semiconductors with strong SOI.

In most lower-dimensional semiconductor structures, the electric fields contributing to SOI have two important contributions: (i) a so-called Dresselhaus field resulting from the lack of inversion symmetry of the crystal structure and (ii) a Rashba field due to asymmetries in the applied confining potential. Although the underlying mechanisms are thus well understood, it still remains a challenge to determine the absolute and relative strength of both contributions in a given sample~\cite{Meier:natphys,Sasaki:natnano}.

Investigating the DC Josephson current through a superconductor-semiconductor-superconductor junction in the presence of an applied magnetic field has been proposed as a way to acquire information about SOI in the semiconductor~\cite{PhysRevB.82.125305,1742-6596-568-5-052035,0295-5075-108-4-47009}. Indeed, SOI can make the current depend anisotropically on the field~\cite{0295-5075-108-4-47009} or produce an anomalous supercurrent (a current at zero phase difference)~\cite{doi:10.7566/JPSJ.82.054703, PhysRevB.89.195407, PhysRevB.82.125305, doi:10.1143/JPSJ.80.124708}. These effects depend on the orientiation and type (Rashba or Dresselhaus) of the SOI and as such could therefore be used to determine or parametrize the SOI in a given sample~\cite{yacoby:arxiv}.

Previous models produced (semi-)analytic results for the Josephson current as a function of SOI parameters, e.g.\ for strictly one-dimensional wires~\cite{0295-5075-108-4-47009,doi:10.7566/JPSJ.82.054703}, for quasi-one-dimensional systems~\cite{krive:lowtemp,krive:prb}, as well as truly two-dimensional junctions~\cite{PhysRevB.66.052508}, including Rashba SOI, electron-electron interactions, and a Zeeman field (i.e.~no induced vector potential).
The appearance of an anomalous Josephson current was shown to rely on the presence of both SOI and a finite exchange field, its magnitude depending on the angle between the two effective fields.

Realistic systems are however usually more complex: They can be disordered, the two contacts to the superconductors can have different transparancies, the vector potential due to the applied magnetic field can have non-negligible effects, or both Rashba and Dresselhaus SOI can be present. Including all these ingredients makes it very challenging to obtain analytic insights, and usually one has to revert to numerics in order to produce quantitative or qualitative results.
Numerical results based on a Keldysh-Usadel approach were used to study the supercurrent in diffusive junctions, with an asymmetric design or inhomogeneous exchange field~\cite{alidoust1,alidoust2}. There it was found that an anomalous supercurrent can also be caused by device asymmetries.

In this work, we perform a most general analysis by investigating symmetries of the full Hamiltonian describing a two-dimensional superconductor-normal metal-superconductor (SNS) junction.
We relate basic properties of the Josephson current to the absence or presence of certain symmetries in the Hamiltonian, and we identify the ingredients that break these symmetries, including Rashba and Dresselhaus SOI.
Our main result is a clear overview that qualitatively links easily observable properties of the supercurrent and the critical current to the structure of the underlying Hamiltonian.
In contrast to similar analyses in the literature~\cite{PhysRevB.89.195407,PhysRevB.82.125305}, we (i) include disorder and a finite vector potential and (ii) do not restrict our investigation to the anomalous current, but also conclude on the magnetic-field-dependence of the critical current.
We support our symmetry analysis with numerical calculations of the Josephson current based on a perturbative expansion in a weak tunnel coupling between the normal region and the superconducting leads.
Explicit calculations that can be found in the literature---mainly concerning the anomalous supercurrent~\cite{krive:lowtemp,krive:prb,PhysRevB.66.052508,alidoust1}---agree with our results.

%\section{Model}
\begin{figure}[t]
\begin{center}
\includegraphics[width=82mm]{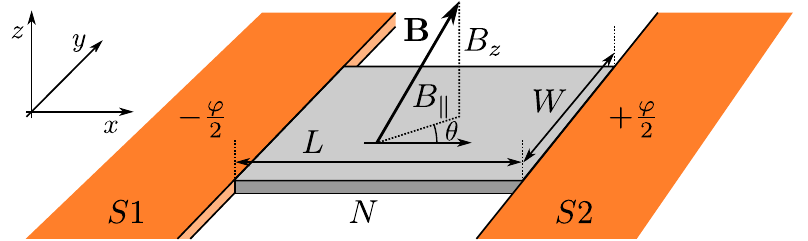}
\caption{(Color online) A two-dimensional SNS junction where a normal metal region of size $W\times L$ is coupled to two two-dimensional superconductors with a phase difference of $\varphi$. We consider the properties of the (critical) supercurrent through the junction in the presence of SOI, arbitrarily oriented magnetic fields, and disorder.}\label{fig:fig1}
\end{center}
\end{figure}
We consider a two-dimensional SNS junction, as shown 
in Fig.~\ref{fig:fig1}. A normal metal region ($N$) with dimensions $W\times L$ 
is coupled to two superconducting leads ($S1$ and $S2$), and we investigate the 
supercurrent through the junction as a function of the phase difference $\varphi$ between the leads. We describe the electrons in the junction with a Bogoliubov-de Gennes Hamiltonian $H = \frac{1}{2} \int d{\bf r} \, \Psi^\dagger {\cal H}\Psi$, using $\Psi = [ \psi_\ua({\bf r}), \psi_\da({\bf r}), \psi^\dagger_\da({\bf r}), -\psi^\dagger_\ua({\bf r})]^T$, where $\psi_{\ua(\da)}({\bf r})$ is the electronic annihilation operator for an electron with spin up(down) at position ${\bf r}$. In this framework, we write
\begin{align}
{\cal H} = {\cal H}_0 + {\cal H}_{\rm SOI} + {\cal H}_{\rm Z} + {\cal H}_{\rm S}.
\end{align}

The first term in ${\cal H}$ reads
\begin{align*}
{\cal H}_0 = \left\{ \frac{p^2}{2m} 
+ \frac{(eB_zy)^2}{2m}
- \mu + V(x,y) \right\}\tau_z + \frac{e}{m} B_z y p_x ,
\end{align*}
where ${\bf p} = -i\hbar(\partial_x, \partial_y)$ is the electronic momentum operator and the Pauli matrices $\boldsymbol \tau$ act in particle-hole space.
This ${\cal H}_0$ thus describes the kinetic energy of the electrons, where we added the effect of a vector potential corresponding to a uniform magnetic field $B_z$ penetrating $N$, using the gauge ${\bf A} = -B_z y\hat x$.
We also included a position-dependent potential $V(x,y)$ to model possible barriers at the SN-interfaces or other effects such as disorder or band bending. The second term describes the spin-orbit interaction of the propagating electrons,
\begin{align*}
{\cal H}_{\rm SOI} {} = {} & {}
\left\{ \frac{\alpha}{\hbar} (- p_y\sigma_x + p_x\sigma_y )+ \frac{\beta}{\hbar}(-p_x\sigma_x + p_y\sigma_y) \right\}\tau_z \nonumber\\
{} & {} + \frac{eB_zy}{\hbar} ( \alpha\sigma_y - \beta\sigma_x),
\end{align*}
the terms proportional to $\alpha$ accounting for the Rashba coupling and those coming with $\beta$ for the Dresselhaus coupling~\cite{fraunhofer_theory:note1}.
${\cal H}_{\rm SOI}$ contains only the linear Dresselhaus terms, but in principle one could also include cubic terms (usually $\propto p_y^2 p_x \sigma_x$, $p_x^2 p_y \sigma_y$), which become important when the thickness of the junction in the $z$-direction is non-negligible.
However, these cubic terms transform identically to the linear Dresselhaus terms under all symmetry operations we consider in this work~\cite{fraunhofer_theory:sup}, and we therefore do not write them explicitly here.
The Zeeman splitting of the electronic spin states is described by
\begin{align*}
{\cal H}_{\rm Z} = \frac{1}{2}g\mu_{\rm B}{\bf B}\cdot\boldsymbol\sigma,
\end{align*}
$g$ being the effective $g$-factor. Finally, the superconductivity in S1 and S2 is modeled by the $s$-wave pairing term
\begin{align*}
{\cal H}_{\rm S} = {} & {} \Delta \Theta(x-\tfrac{1}{2}L) \left[ \cos \tfrac{\varphi}{2}\tau_x + \sin\tfrac{\varphi}{2}\tau_y \right] \\
& {} + {} \Delta \Theta(-x-\tfrac{1}{2}L) \left[ \cos \tfrac{\varphi}{2}\tau_x - \sin\tfrac{\varphi}{2}\tau_y \right],
\end{align*}
where $\Delta$ is the magnitude of the pairing potential and $\Theta(x)$ is the Heaviside step-function. We note that, in this description, the parameters $m$, $\mu$, $\alpha$, $\beta$, and $g$ can be position-dependent (i.e.\ can effectively vary from the normal to the superconducting regions). In this case, a Hermitian Hamiltonian can be obtained by symmetrization of terms containing momentum operators and position-dependent parameters, e.g.\ $\alpha(x)\partial_x \to \tfrac{1}{2}[\alpha(x)\partial_x + \partial_x \alpha(x)]$. Assuming all variations to be symmetric under $x\to -x$, this does not make a difference for the arguments to follow.

With $k_{\rm B} = 1$, the free energy of the junction follows as
\begin{equation}
F = -T\ln\mathrm{Tr}\{ e^{-H/T}\},\label{eq:freeen}
\end{equation}
up to a phase-independent constant. The supercurrent through the junction is then calculated from the thermodynamic relation
\begin{equation}
I_s(\varphi) = \frac{2e}{\hbar} \frac{\partial F}{\partial \varphi}.
\end{equation}
The critical current $I_{c\pm}$ in both directions follows as
\begin{equation}
I_{c+}= \max_\varphi I_s(\varphi) \quad\text{and}\quad I_{c-} = \min_\varphi I_s(\varphi),
\end{equation}
and the anomalous Josephson current is $I_{\rm an} = I_s(\varphi=0)$.

%\section{Symmetry properties}
We now proceed with a general investigation of the symmetry properties of $H$ and their implications for the supercurrent through the junction. Our investigation is based on the fact that any two Hamiltonians $H$ and $H'$ have identical spectra if and only if there exists a unitary or antiunitary transformation $U$ such that $H = UH'U^\dagger$. In that case, it follows straightforwardly from (\ref{eq:freeen}) that the transformation does not affect the free energy, $F'=F$, which implies a relation between $I_s'$ and $I_s$. Investigating transformations between specific pairs of $H$ and $H'$ thus allows us to find symmetries that the supercurrent has to possess, as well as necessary requirements for asymmetries to be present in the supercurrent.

%\subsection{Anomalous Josephson current}
We first focus on the symmetries related to the anomalous Josephson current. 
In order to have a finite supercurrent at zero phase difference, $I_s(\varphi = 
0) \neq 0$, it is required that the free energy is \textit{not} symmetric under 
the transformation $\varphi \to -\varphi$. Indeed, if $F(\varphi) = F(-\varphi)$ 
then it follows that $I_s(\varphi) = -I_s(-\varphi)$, which in turn implies that 
the anomalous Josephson current vanishes. Thus, investigating in which cases there exist $U$ such that $UH(\varphi)U^\dagger = H(-\varphi)$ allows 
us to determine necessary conditions for an anomalous supercurrent to occur.

Swapping the sign of $\varphi$ in the Hamiltonian can be achieved by (i) the $x$-parity operation ${\cal P}_x$, which effectively interchanges the two superconductors $S1$ and $S2$, and (ii) time-reversal $T = i\sigma_yK$. For a minimal Fraunhofer interference model (for now without SOI, an in-plane magnetic field, and disorder) we use
\begin{align*}
{\cal H}_{{\rm min}} = {} & {} \left\{ \frac{p^2}{2m} + \frac{(eB_zy)^2}{2m}- \mu \right\}\tau_z + \frac{e}{m} B_z y p_x \nonumber\\
{} & {} + \frac{1}{2}g\mu_{\rm B}B_z\sigma_z+ {\cal H}_{\rm S}.
\end{align*}
The parity operation ${\cal P}_x$ changes the sign of the third term in ${\cal H}_{{\rm min}}$, and time-reversal changes the sign of both terms proportional to $B_z$. Therefore, we can construct four symmetry operators that effect $UH(\varphi)U^\dagger = H(-\varphi)$ in this minimal setup: (i) ${\cal P}_x{\cal P}_y$, (ii) $\sigma_z{\cal P}_x{\cal P}_y$, (iii) $\sigma_x {\cal P}_y T$, and (iv) $\sigma_y{\cal P}_yT$. As long as at least one of these symmetries is retained, one has $I_s(\varphi) = -I_s(-\varphi)$ which implies $I_{\rm an} = 0$.

\begin{table}[b]
\begin{tabular}{ll}
\multicolumn{2}{c}{$UH(\varphi)U^\dagger = H(-\varphi)$}\\
\colrule
$U$ & broken by\\
\colrule
$\mathcal{P}_y\mathcal{P}_x$ & $\alpha$, $\beta $, $V_x$, $V_y$\\
$\sigma_z\mathcal{P}_y \mathcal{P}_x$ & $B_x$, $B_y$, $V_x$, $V_y$\\
$\sigma_x \mathcal{P}_y T$ & $B_x$, $\alpha$, $V_y$\\
$\sigma_y \mathcal{P}_y T$ & $B_y$, $\beta$, $V_y$
\end{tabular}
\caption{Operators $U$ that effect $UH(\varphi)U^\dagger = H(-\varphi)$ (in the presence of a perpendicular magnetic field $B_z$) and possible extra ingredients in the Hamiltonian that would break these symmetries. $V_{x(y)}$ indicates the presence of a potential $V(x,y)$ that is asymmetric under ${\cal P}_{x(y)}$.}
\label{tab:table1}
\end{table}
Going now to the full model Hamiltonian, where SOI, a finite in-plane magnetic field, and an asymmetric potential can be present, these symmetries can be broken, allowing for a finite anomalous Josephson current. In Table \ref{tab:table1} we list the four symmetry transformations and the extra ingredients in the Hamiltonian that break them. Here $V_{x(y)}$ signals that the potential $V(x,y)$ is asymmetric under ${\cal P}_{x(y)}$. The presence of Rashba(Dresselhaus) SOI is indicated by $\alpha$($\beta$), and a finite in-plane magnetic field along $\hat x$($\hat y$) by $B_{x(y)}$ \cite{fraunhofer_theory:note2}. We can now straightforwardly identify combinations of ingredients that allow for a finite anomalous supercurrent. For instance (i) Rashba SOI and a finite $B_y$, or (ii) Dresselhaus SOI and a finite $B_x$, or (iii) the mere presence of $V_y$. In a clean sample, measuring $I_{\rm an}$ while rotating the in-plane magnetic field thus reveals information about the presence or absence of Rashba and Dresselhaus coupling separately.

%\subsection{Fraunhofer pattern}
Similarly, we can investigate symmetries related to the magnetic-field-dependence of the critical currents $I_{c\pm}(B_z)$. The existence of operations $U$ yielding $UH(B_z,\varphi)U^\dagger = H(-B_z,\varphi)$ implies that $I_s(B_z,\varphi)=I_s(-B_z,\varphi)$ so that the critical current must be symmetric in $B_z$, i.e.~$I_{c\pm}(B_z) = I_{c\pm}(-B_z)$.
For the minimal setup, we can identify four such operations,
which are listed in the left part of Table~\ref{tab:table2} together with the extra terms that would break the corresponding symmetries.
\begin{table}[t]
\begin{tabular}{lll|lll}
\multicolumn{3}{c|}{$UH(B_z,\varphi)U^\dagger = H(-B_z,\varphi)$} &
\multicolumn{3}{c}{$UH(B_z,\varphi)U^\dagger = H(-B_z,-\varphi)$} \\
\colrule
\hspace{5pt} & $U$ & broken by & \hspace{1em} & $U$ & broken by\\
\colrule
 & $\sigma_x {\cal P}_y$ & $B_y$, $\alpha$, $V_y$ &
 & $T$ & $B_x$, $B_y$
 \\
 & $\sigma_y {\cal P}_y$ & $B_x$, $\beta$, $V_y$ &
 & $\sigma_zT$ & $\alpha$, $\beta$
 \\
 & ${\cal P}_x{\cal P}_yT$ & $B_x$, $B_y$, $\alpha$, $\beta$, &
 & $\sigma_x \mathcal{P}_x$ & $B_y$, $\beta$, $V_x$
 \\
 &  & $V_x$, $V_y$ &
 & $\sigma_y \mathcal{P}_x$ & $B_x$, $\alpha$, $V_x$
 \\
 & $\sigma_z{\cal P}_x{\cal P}_yT$ & $V_x$, $V_y$ &
 & &
\end{tabular}
\caption{Left: Operators $U$ that yield $UH(B_z,\varphi)U^\dagger = H(-B_z,\varphi)$ and possible extra ingredients in the Hamiltonian that would break these symmetries. Right: The same, but for the transformation $UH(B_z,\varphi)U^\dagger = H(-B_z,-\varphi)$.}
\label{tab:table2}
\end{table}
We see that in the absence of disorder this symmetry will always be present. However, with a finite $V_x$ (e.g.~due to asymmetric barriers at the SN interfaces) an asymmetry will develop for (i) Rashba SOI and a finite $B_x$ or (ii) Dresselhaus SOI and a finite $B_y$. (Note that the combinations are opposite from the ones giving rise to an anomalous current.) This symmetry thus presents a second independent way to obtain information about the presence or absence of Rashba and Dresselhaus coupling.
 
Another symmetry that can be present in the pattern of critical currents is reflection symmetry with respect to the axis $I_c = 0$, i.e. the maximum and minimum Josephson current are equal, $I_{c-}(B_z) = -I_{c+}(B_z)$. This symmetry is guaranteed to be present if there exists a $U$ such that $UH(B_z,\varphi)U^\dagger = H(B_z,-\varphi)$. This is the same transformation as the one connected to the vanishing of the anomalous Josephson current, and the results presented in Table~\ref{tab:table1} apply again.

Finally, we investigate operations that yield $UH(B_z,\varphi)U^\dagger = H(-B_z,-\varphi)$. The existence of such a symmetry would imply $I_s(B_z,\varphi) = -I_s(-B_z,-\varphi)$ and thus $I_{c+}(B_z) = -I_{c-}(-B_z)$, meaning that the interference pattern of critial currents would be inversion symmetric through the point ($B_z=0$, $I_c=0$). We list the four relevant symmetry operations
in the right part of Table~\ref{tab:table2}, again indicating which extra terms in the Hamiltonian would break the corresponding symmetries.

This concludes our overview of the main symmetry properties of $I_{c\pm}(B_z,\varphi)$ and $I_{\rm an}(\varphi)$. With the help of the Tables presented here, easily observable quantities (the anomalous supercurrent and the basic symmetries of the pattern of critical currents) can be directly related to the direction of the magnetic field, the presence or absence of different types of spin-orbit coupling, and asymmetries in the potential.

%\section{Numerical investigation}
To support the above results, we present numerical calculations of the supercurrent in the SNS geometry. Our calculations are based on a perturbative expansion of the free energy of the normal region, assuming weak coupling between the normal region and the superconductors~\cite{doi:10.1143/JPSJ.68.954}.
% The corrections to the unperturbed free energy $F_0$ follow as~\cite{adg:book}
% \begin{align}
% F = F_0 - T(\langle {\cal S} \rangle_{\rm con} - 1),
% \end{align}
% in terms of all fully connected correlation functions that can be constructed from an expansion of the ${\cal S}$-matrix
% \begin{align}
% {\cal S}(\tau) = T_\tau \exp \bigg\{ -\int_0^\tau d\tau'\, [H_{T1}(\tau') + H_{T2}(\tau') ] \bigg\},
% \end{align}
% where $T_\tau$ is the time-ordering operator for the imaginary time $\tau$.
After integrating out the superconductors, we find to leading order in the coupling~\cite{fraunhofer_theory:sup}
%and integrating out the degrees of freedom of the superconductors, the $\varphi$-dependent corrections to $F_0$ yield
\begin{align}
I_s(\varphi) = & -{\rm Im} \bigg[ e^{-i\varphi} \frac{4eT}{\hbar}\sum_{n,\alpha,\beta}\int_{-\tfrac{W}{2}}^{\tfrac{W}{2}}dy_1dy_2 \,
\frac{(\kappa W \Delta)^{2}}{\Delta^{2}+\omega_{n}^{2}} \nonumber \\
 & \times \alpha\beta\, {\cal G}_{\beta\alpha}^{RL}(y_2,y_1;i\omega_{n}){\cal G}^{RL}_{\bar \beta\bar \alpha}(y_2,y_1;-i\omega_{n}) \bigg],\label{eq:supcur}
\end{align}
where $\kappa$ parametrizes the coupling, $\alpha,\beta = \pm 1$ denote the two spin directions, $\omega_n = (2n\pi +1)T/\hbar$ are the fermionic Matsubara frequencies, and $\nu_{\rm sc}$ is the normal-state density of states of the superconductors. The Green functions ${\cal G}^{RL}$ are related to the amplitudes for propagation of an electron in the normal region from the left contact to the right contact,
\begin{align}
{\cal G}_{\beta\alpha}^{RL}&(y_2,y_1;i\omega) \\
 & = -\frac{1}{\hbar}\int_0^{\hbar/T}\!\!\!\! d\tau\,e^{i\omega\tau}\langle T_\tau \psi_\beta^\dagger (\tfrac{L}{2},y_2;\tau)\psi_\alpha (-\tfrac{L}{2},y_1;0)\rangle. \nonumber
\end{align}
We emphasize that Eq.~(\ref{eq:supcur}) only evaluates the lowest Fourier component of the supercurrent, and therefore always results in $I_s(\varphi) = I_c \sin(\varphi - \varphi_0)$ and $I_{c+} = I_{c-}$. Asymmetries in $I_{c\pm}$ only appear in higher-order corrections to $I_s(\varphi)$, as we will show below.

\begin{figure}[t]
\begin{center}
\includegraphics[width=82mm]{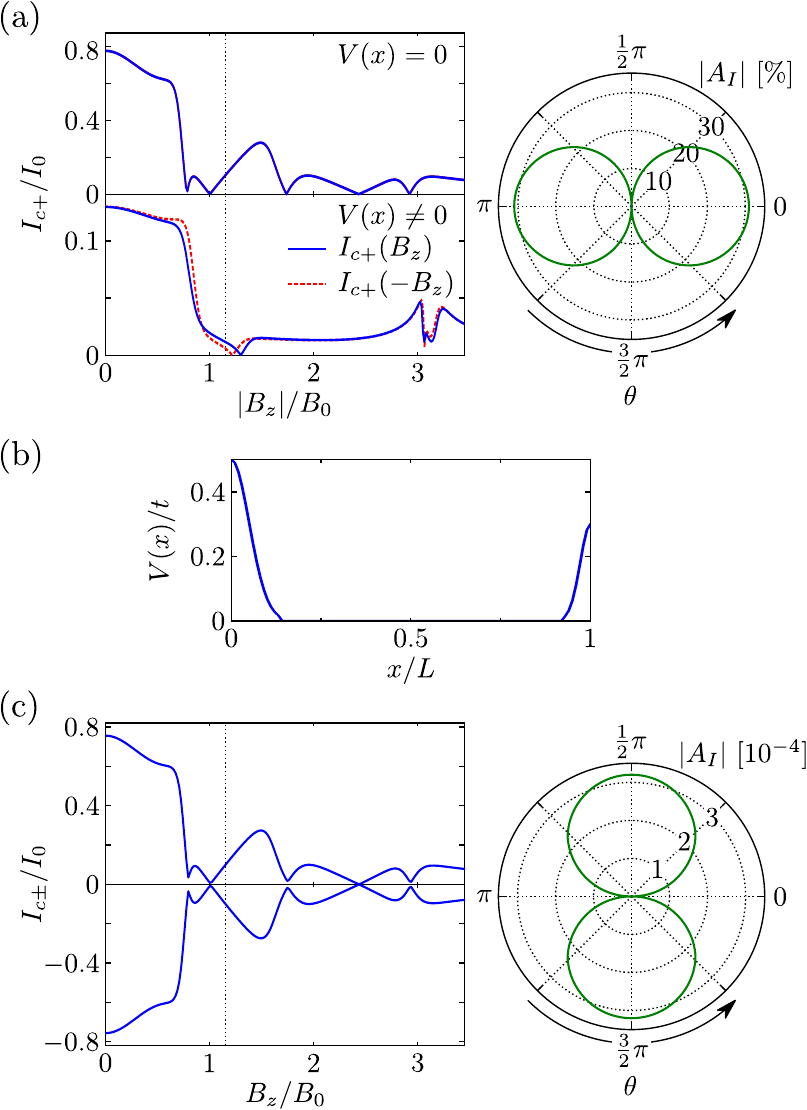}
\caption{(Color online)
Asymmetries in the pattern of critical currents.
(a)
Left: $I_{c+}(B_z)$ and $I_{c+}(-B_z)$ for $B_x = 200$~mT and $B_y = 0$, without (upper plot) and with (lower plot) the symmetry-breaking potential $V(x)$ shown in (b).
Current is plotted in units of $I_0 = \Delta e / h$ and $B_z$ is normalized to the field for which one superconducting flux quantum penetrates the normal area $B_0=h/2eA$.
Right: Dependence of the asymmetry in $I_{c+}(\pm B_z)$ on the direction $\theta$ of the in-plane magnetic field, at $B_z = 80$~mT, indicated by the dotted line in the left plot.
(c)
Left: $I_{c\pm}(B_z)$ for $\theta = \frac{1}{2}\pi$ (to fourth order in $\kappa$), with $V(x)=0$.
Right: Dependence of the asymmetry in $I_{c\pm}(B_z)$ on $\theta$ at $B_z = 80$ mT.
}\label{fig:fig2}
\end{center}
\end{figure}
For our numerical calculations, we assume a normal region with width $W = 99$~nm and length $L=300$~nm~\cite{footnote3}, and we discretize the Hamiltonian $H_{\rm N}$ on a lattice with lattice constant $a = 3$~nm, resulting in a hopping matrix element $t = \hbar^2 / 2ma^2 = 163$ meV (assuming an effective mass of $m = 0.026\, m_e$).
We set the Fermi wavelength to $\lambda_{\rm F} = 20$~nm, which corresponds to $\mu_N = 0.89\,t$, and use a $g$-factor of $g = -10$, yielding a ``Zeeman length'' $l_{\rm Z} = 2\pi \hbar v_{\rm F} / |g|\mu_{\rm B} B \approx 10~\mu$m for $B=1$~T.
The superconducting pairing potential in the leads is set to $\Delta = 0.1$~meV, so that the coherence length $\xi = \hbar v_{\rm F} / \pi \Delta \approx 3~\mu$m and we are in the short-junction limit.
Further, we use a temperature $T = 100$ mK and NS coupling parameter $\kappa = 3$~meV.
The required Green functions ${\cal G}^{RL}$ can be found from solving for elements of $[i\hbar\omega-H_N]^{-1}$, and the supercurrent through the junction then follows straightforwardly from (\ref{eq:supcur}), where the integrals over $y_{1,2}$ are replaced by sums over lattice sites.

\begin{figure}[t]
\begin{center}
\includegraphics[width=82mm]{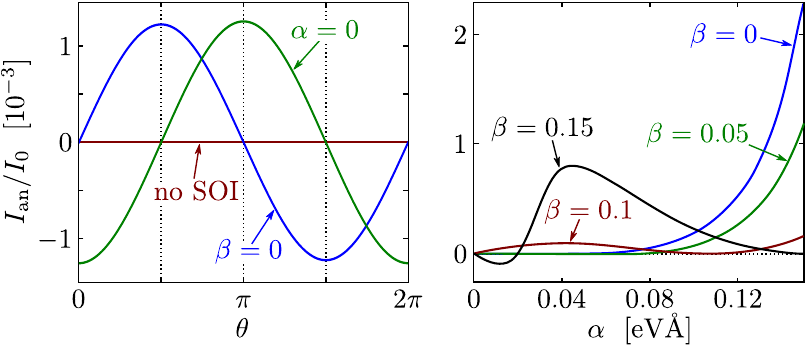}
\caption{(Color online)
Asymmetries in the anomalous supercurrent.
Left: $I_{\rm an}$ as a function of the orientation of the in-plane field, for different values of $\alpha$ and $\beta$ and with $V(x,y)=0$.
Right: $I_{\rm an}$ as a function of small $\alpha$ with $\theta = \frac{1}{2}\pi$, for four different values of $\beta$.
}\label{fig:fig3}
\end{center}
\end{figure}

The results are presented in Figs.~\ref{fig:fig2} and \ref{fig:fig3}.
In Fig.~\ref{fig:fig2}a we first investigate the symmetry $I_{c+}(B_z) = I_{c+}(-B_z)$.
In the left panel we plot $I_{c+}(B_z)$ (solid blue) and $I_{c+}(-B_z)$ (dashed red) for $B_x = 200$~mT and $B_y = 0$.
We use the spin-orbit parameters $\alpha = 0.921$ eV\AA{} and $\beta = 0$, which corresponds to a spin-orbit length comparable to the junction size, $l_{\rm so} = \pi \hbar^2 / \alpha m \approx 100$~nm.
The current is plotted in units of $\Delta e / h$ and the field $B_z$ is normalized by the field corresponding to one (superconducting) flux quantum penetrating the normal region $B_0 = h/2eA$, where $A = WL$.
From Table~\ref{tab:table2} we see that the critical current is expected to be symmetric as long as $V(x,y) = V(-x,-y)$.
Indeed, with $V(x,y) = 0$ (upper part) the critical current is equal for $\pm B_z$ and the two curves fall on top of each other.
In the lower panel we show the critical current when the symmetry of $V(x,y)$ is broken by including the $x$-dependent potential $V(x)$ shown in Fig.~\ref{fig:fig2}b.
As expected, the two critical currents $I_{c+}(B_z)$ and $I_{c-}(B_z)$ now are different.
In the right panel of Fig.~\ref{fig:fig2}a we illustrate the dependence of this asymmetry on the orientation of the in-plane magnetic field.
For this plot we fix $B_z = 80$~mT $\approx 1.15\,B_0$, and we plot as a measure for the asymmetry $|A_I| \equiv |[I_{c+}(B_z) - I_{c+}(-B_z)] / [I_{c+}(B_z) + I_{c+}(-B_z)]|$, as a function of the angle $\theta$ between the in-plane magnetic field ${\bf B}_\parallel$ and the $x$-axis (using $B_\parallel = 200$ mT).
The disorder potential $V(x)$ is again the one shown in Fig.~\ref{fig:fig2}b, and we find that the asymmetry is maximal for ${\bf B}_\parallel \parallel \hat x$ and vanishes for ${\bf B}_\parallel \parallel \hat y$, as expected from Table~\ref{tab:table2}.

In Fig.~\ref{fig:fig2}c we focus on the symmetry $I_{c+}(B_z)=-I_{c-}(B_z)$.
As explained before, Eq.~(\ref{eq:supcur}) only produces the lowest Fourier component of $I_s(\varphi)$, so its minimum and maximum values have to have equal magnitudes.
To make asymmetries in $I_{c\pm}$ visible, we thus add the next-order correction (fourth order in $\kappa$) to $I_s(\varphi)$, see~\cite{fraunhofer_theory:sup} for the details.
The left panel shows the resulting $I_{c+}(B_z)$ and $-I_{c-}(B_z)$ using the same set of parameters as above, with $V(x,y)=0$ and ${\bf B}_\parallel \parallel \hat y$.
We see that (i) the pattern of critical currents looks very similar to the second-order result shown in the upper left panel of Fig.~\ref{fig:fig2}a and (ii) any asymmetry between $I_{c+}$ and $I_{c-}$, if present at all, is small.
Both these observations are consistent with the fact that all deviations from the results presented in the top left plot of Fig.~\ref{fig:fig2}a are due to small higher-order corrections.
To investigate the symmetries as predicted by Table~\ref{tab:table1} in more detail, we show in the right panel $A_I = [I_{c+} - |I_{c-}|]/[I_{c+} + |I_{c-}|]$ (a measure for the asymmetry) as a function of the in-plane angle of the magnetic field.
As expected, the critical current is symmetric $I_{c+} = - I_{c-}$ when $B_y = 0$ (i.e.~$\theta = 0,\pi$) and asymmetric for all other orientations of the in-plane field.

Finally, in Fig.~\ref{fig:fig3} we investigate the anomalous Josephson current.
In the left panel we plot $I_{\rm an}$ as a function of $\theta$,
where we have set $B_z/B_0 = 1.80$, $B_\parallel = 200$~mT, and $V(x,y) = 0$.
The three curves correspond to $\alpha = 0.921$~eV\AA{}, $\beta = 0$ (blue), $\alpha = 0$, $\beta = 0.921$ eV\AA{} (green), and $\alpha = \beta = 0$ (red).
The current indeed vanishes at all points predicted by Table~\ref{tab:table1} and is non-zero everywhere else.
In the right panel we investigate the symmetry-breaking for small $\alpha$ in more detail:
We plot $I_{\rm an}$ with $\theta = \frac{1}{2}\pi$ for $\alpha$ ranging from zero to $0.15$ eV\AA{}, which corresponds to $\pi \hbar^2 / \alpha m \gtrsim 600$~nm.
Using four different values of $\beta$, being $\beta = 0$ (blue curve), $\beta = 0.05$~eV\AA{} (green), $\beta = 0.1$~eV\AA{} (red), and $\beta = 0.15$~eV\AA{} (black),
we see that the qualitative behavior of $I_{\rm an}(\alpha)$ at small $\alpha$ can vary strongly with the choice of other parameters.

To summarize, we presented a general symmetry analysis of a model Hamiltonian describing a two-dimensional SNS junction, including an arbitrarily oriented magnetic field, both Rashba and Dresselhaus spin-orbit interaction, and disorder and other structural asymmetries. We related basic properties of the anomalous current and the critical currents to the absence or presence of specific ingredients in the Hamiltonian, thereby providing a qualitative connection between easily measurable quantities and the relative strength of the different underlying mechanisms. We supported our analysis with numerical calculations of the Josephson current, agreeing with the qualitative predictions we made.

We gratefully acknowledge stimulating discussions with C.~M.~Marcus.
The Center for Quantum Devices is funded by the Danish National Research Foundation. The research was supported by The Danish Council for Independent Research|Natural Sciences. F.N.\ acknowledges support from the European Community through the Marie Curie Fellowship, grant agreement No 659653.

\newpage
\mbox{}
\newpage
\widetext
\begin{center}
\textbf{\large Supplementary Information: Effects of spin-orbit coupling and spatial symmetries\\ on the Josephson current in SNS junctions}
\end{center}
%%%%%%%%%% Merge with supplemental materials %%%%%%%%%%
%%%%%%%%%% Prefix a "S" to all equations, figures, tables and reset the counter %%%%%%%%%%
\setcounter{equation}{0}
\setcounter{figure}{0}
\setcounter{table}{0}
\setcounter{page}{1}
\makeatletter

\section{Symmetry transformations fulfilling $UH(\varphi)U^\dagger = H(-\varphi)$}

\begin{table}[b]
\begin{tabular}{ll}
$U$ & broken by\\
\colrule
$T$ & $A_x$, $B_x$, $B_y$, $B_z$, $B_zy\sigma_x$, $B_zy\sigma_y$\\
$\sigma_x T$ & $A_x$, $B_x$, $p_x\sigma_y$, $p_y\sigma_y$, $B_zy\sigma_x$\\
$\sigma_y T$ & $A_x$, $B_y$, $p_y\sigma_x$, $p_x\sigma_x$, $B_zy\sigma_y$\\
$\sigma_z T$ & $A_x$, $B_z$, $p_x\sigma_y$, $p_y\sigma_x$, $p_x\sigma_x$, $p_y\sigma_y$\\
$\mathcal{P}_y T$ & $B_x$, $B_y$, $B_z$, $p_y\sigma_x$, $p_y\sigma_y$, $V_y$\\
$\mathcal{P}_y \sigma_x T$ & $B_x$, $p_x\sigma_y$, $p_y\sigma_x$, $V_y$, $B_zy\sigma_y$\\
$\mathcal{P}_y \sigma_y T$ & $B_y$, $p_x\sigma_x$, $p_y\sigma_y$, $V_y$, $B_zy\sigma_x$\\
$\mathcal{P}_y \sigma_z T$ & $B_z$, $p_x\sigma_y$, $p_x\sigma_x$, $V_y$, $B_zy\sigma_x$, $B_zy\sigma_y$\\
$\mathcal{P}_x$ & $A_x$, $p_x\sigma_y$, $p_x\sigma_x$, $V_x$\\
$\mathcal{P}_x \sigma_x$ & $A_x$, $B_y$, $B_z$, $p_x\sigma_x$, $p_y\sigma_y$, $V_x$, $B_zy\sigma_y$\\
$\mathcal{P}_x \sigma_y$ & $A_x$, $B_x$, $B_z$, $p_x\sigma_y$, $p_y\sigma_x$, $V_x$, $B_zy\sigma_x$\\
$\mathcal{P}_x \sigma_z$ & $A_x$, $B_x$, $B_y$, $p_y\sigma_x$, $p_y\sigma_y$, $V_x$, $B_zy\sigma_x$, $B_zy\sigma_y$\\
$\mathcal{P}_x\mathcal{P}_y$ & $p_x\sigma_y$, $p_y\sigma_x$, $p_x\sigma_x$, $p_y\sigma_y$, $V_x$, $V_y$, $B_zy\sigma_x$, $B_zy\sigma_y$\\
$\mathcal{P}_x\mathcal{P}_y\sigma_x$ & $B_y$, $B_z$, $p_y\sigma_x$, $p_x\sigma_x$, $V_x$, $V_y$, $B_zy\sigma_x$\\
$\mathcal{P}_x\mathcal{P}_y\sigma_y$ & $B_x$, $B_z$, $p_x\sigma_y$, $p_y\sigma_y$, $V_x$, $V_y$, $B_zy\sigma_y$\\
$\mathcal{P}_x\mathcal{P}_y\sigma_z$ & $B_x$, $B_y$, $V_x$, $V_y$\\
\end{tabular}
\caption{Operators $U$ that effect $UH(\varphi)U^\dagger = H(-\varphi)$ and possible extra ingredients in the Hamiltonian that would break these symmetries. The asymmetry $A_x$ indicates a finite vector potential ${\bf A} = (-B_zy, 0)$. We also indicate separately the effect of all six spin-orbit terms, including those that are only present when a field $B_z$ is applied.}
\label{tab:tab1}
\end{table}
\noindent If the $z$-component of the magnetic field is not part of the minimal Hamiltonian,
\begin{align}
{\cal H}_{{\rm min}} = \left\{ \frac{p^2}{2m} - \mu\right\}\tau_z + {\cal H}_{\rm S},
\label{eq:hnmin}
\end{align}
we can in fact find 16 symmetry operations that enact $UH_{\rm min}(\varphi)U^\dagger = H_{\rm min}(-\varphi)$, which are $\sigma_n T$, ${\cal P}_y \sigma_n T$, ${\cal P}_x \sigma_n$, and ${\cal P}_x{\cal P}_y\sigma_n$, where $n \in \{0,x,y,z\}$ with $\sigma_0$ denoting the unity matrix in spin space. Treating $B_z$ as one of the possible symmetry breaking ingredients, we then arrive at the overview given in Table~\ref{tab:tab1}. To distinguish the corresponding symmetries, we treat the Zeeman splitting due to $B_z$ and the vector potential $A_x = -B_z y$ associated to $B_z$ as two different ingredients, the presence of the latter is indicated by $A_x$.

We also split the two standard spin-orbit terms (Rashba and Dresselhaus) into their constituents $p_x\sigma_x$, $p_y\sigma_x$, $p_x\sigma_y$, $p_y\sigma_y$, $B_zy\sigma_y$, and $B_zy\sigma_x$ (in the absence of $B_z$ the last two terms are of course neither present).
In some cases one might be interested in the effect of only a subset of these terms.
In case the $x$-direction in our description does not correspond to the $[100]$ crystallographic direction, then the spin-orbit Hamiltonian becomes
\begin{align}
{\cal H}_{\rm SOI} = {} & {} \left\{ \frac{\alpha}{\hbar} (- p_y\sigma_x + p_x\sigma_y )+ \frac{\beta}{\hbar} (-p_x\sigma_x + p_y\sigma_y)\cos 2\chi + \frac{\beta}{\hbar}(p_y\sigma_x + p_x\sigma_y)\sin 2\chi \right\}\tau_z
\nonumber \\
{} & {} + \frac{eB_zy}{\hbar} \big\{ (\alpha + \beta \sin 2\chi) \sigma_y - (\beta \cos 2\chi) \sigma_x \big\},
\end{align}
where $\chi$ is the angle between the $x$-axis and the $[100]$ direction (assuming a zinc-blende crystal structure). We see that for special sample orientations and special values of $\alpha$ and $\beta$ some terms could be absent. For instance, with $\chi = \frac{1}{4}\pi$ and $\alpha = \beta$ we have
\begin{align}
{\cal H}_{\rm SOI} = \frac{2 \alpha}{\hbar} ( p_x\tau_z + eB_zy) \sigma_y,
\end{align}
in which case only one spin-orbit direction is of interest.

\section{Cubic Dresselhaus terms}

In the presence of the vector potential ${\bf A} = (-B_z y, 0)$, the two cubic Dresselhaus terms, which are usually described by the Hamiltonian $H_{\rm SOI,3} = \tilde\beta ( p_y^2p_x \sigma_x - p_x^2p_y\sigma_y)$, have to be properly symmetrized and read
\begin{align}
H_{\rm SOI,3} =
\frac{\tilde{\beta}}{4}\big(\{ \{ p_{x}+eB_{z}y,p_{y}\} ,p_{y}\} \sigma_{x}-\{ \{ p_{y},p_{x}+eB_{z}y\} ,p_{x}+eB_{z}y\} \sigma_{y}\big).
\end{align}
This yields explicitly
\begin{align}
{\cal H}_{\rm SOI,3} = 
\tilde{\beta}\Big( {} & {} \left[p_{x}p_{y}^{2}\tau_z+eB_{z}yp_{y}^{2}-i\hbar eB_{z}p_{y}\right]\sigma_{x} \nonumber \\
{} & {} 
- \left[p_{x}^{2}p_{y}\tau_z+2eB_{z}yp_{x}p_{y}+(eB_{z}y)^{2}p_{y}\tau_z-i\hbar eB_{z}p_{x}-i\hbar(eB_{z})^{2}y\tau_z \right]\sigma_{y}\Big),
\end{align}
where we now wrote the Hamiltonian in the required Bogoliubov-de Gennes form.

An inspection of the eight terms in this Hamiltonian shows that all of them are invariant under the symmetry operations ${\cal P}_x{\cal P}_y$ and $\sigma_y{\cal P}_yT$, and antisymmetric under $\sigma_z{\cal P}_x{\cal P}_y$ and $\sigma_x {\cal P}_yT$.
They would thus appear in Table I at exactly the same place as the linear Dresselhaus terms indicated by $\beta$ in the Table.
Similarly we find for Table II that, after reversing $B_z$ the operations $\sigma_x{\cal P}_y$, $\sigma_z{\cal P}_x{\cal P}_yT$, $T$, and $\sigma_y{\cal P}_x$ bring all terms back to their original form, and the operations $\sigma_y{\cal P}_y$, ${\cal P}_x{\cal P}_yT$, $\sigma_zT$, and $\sigma_x{\cal P}_x$ yield a minus sign.
This behavior is again identical to that of the linear terms with $\beta$ listed in Table II.
Inclusion of the two ``standard'' cubic Dresselhaus terms thus does not make a difference for the symmetry analysis presented in the main text.

\section{Perturbative calculation of $I_s(\varphi)$}

\noindent The current in the ground state is given by
\begin{equation}
I_s(\varphi) = \frac{2e}{\hbar}\frac{\partial F}{\partial\varphi},
\end{equation}
where $\varphi$ is the phase difference between the two superconductors
and $F$ is the free energy, $F = -T \ln {\rm Tr}\{e^{-H/T}\}$.

We assume that only the free energy of the normal region $F_{\rm N}$ depends significantly on the phase difference between the two superconductors.
We thus split the system in three parts, and write separate Hamiltonians for the normal region and for the two supercondcutors.

For the normal part we use
\begin{align}
H_{\rm N} = \int d{\bf r}\, \boldsymbol\psi^\dagger ({\bf r}) \bigg\{ {} & {}
-\frac{\hbar^2\nabla^2}{2m} - \mu_{\rm N} - i\frac{e\hbar}{m}B_zy\frac{\partial}{\partial x} + \frac{(eB_zy)^2}{2m} + V({\bf r}) 
+ \frac{1}{2}g\mu_{\rm B} {\bf B}\cdot \boldsymbol\sigma \nonumber\\
{} & {} + \alpha \left( i\frac{\partial}{\partial y}\sigma_x - i\frac{\partial}{\partial x}\sigma_y + \frac{eB_zy}{\hbar}\sigma_y \right)
+ \beta \left( i\frac{\partial}{\partial x}\sigma_x - i\frac{\partial}{\partial y}\sigma_y - \frac{eB_zy}{\hbar} \sigma_x \right) \bigg\} \boldsymbol\psi ({\bf r}),
\end{align}
with $\boldsymbol \psi ({\bf r}) = [ \psi_\ua ({\bf r}), \psi_\da ({\bf r})]^T$, where $\psi_{\ua(\da)}({\bf r})$ is the electronic annihilation operator for an electron with spin up(down) at point ${\bf r}$.

The two superconductors are described by the Hamiltonians
\begin{align}
H_{{\rm S}n} = \frac{1}{2} \int d{\bf r}\, \boldsymbol\Psi_n^\dagger ({\bf r}) \bigg\{
\left( -\frac{\hbar^2\nabla^2}{2m_{{\rm S}n}} - \mu_{{\rm S}n} \right) \tau_z
+ \Delta \left( \cos \varphi_n \tau_x + \sin \varphi_n \tau_y \right) \bigg\} \boldsymbol \Psi_n ({\bf r}),
\label{eq:hams}
\end{align}
where $n \in \{ L,R \}$. We use the notation $\boldsymbol \Psi_n ({\bf r}) = [ \Psi_{n,\ua}({\bf r}), \Psi_{n,\da}({\bf r}), \Psi^\dagger_{n,\da}({\bf r}), -\Psi^\dagger_{n,\ua}({\bf r})]^T$ where $\Psi_{n,\ua(\da)}({\bf r})$ is the electronic annihilation operator for an electron with spin up(down) at point ${\bf r}$ in superconductor $n$. In (\ref{eq:hams}), $\Delta$ is the magnitude of the order parameter in the two superconductors (assumed to be the same) and $\varphi_n$ is its phase in superconductor $n$. We neglect the effects of the magnetic field and spin-orbit interaction in the two superconductors.

The coupling between the normal region and the superconductors is described by
\begin{align}
H_{\rm C} = \sum_\sigma \int d{\bf r}\, \left[ \gamma_L({\bf r}) \Psi^\dagger_{L,\sigma}({\bf r})\psi_\sigma ({\bf r}) + 
\gamma_R({\bf r}) \Psi^\dagger_{R,\sigma}({\bf r})\psi_\sigma ({\bf r}) + {\rm H.c.} \right],
\end{align}
where $\gamma_n ({\bf r}) = \gamma\,\delta(x-x_n)$ parametrizes the strength of the coupling, with $x_L = -L/2$ and $x_R = L/2$ indicating the positions of the contacts. This yields
\begin{align}
H_{\rm C} = \gamma \sum_\sigma \int dy\, \left[ \Psi^\dagger_{L,\sigma}(x_L,y)\psi_\sigma (x_L,y) + 
\Psi^\dagger_{R,\sigma}(x_R,y)\psi_\sigma (x_R,y) + {\rm H.c.} \right].
\end{align}

The coupling Hamiltonian $H_{\rm C}$ is treated as a small perturbation, from which we define a so-called ${\cal S}$-matrix by
\begin{equation}
{\cal S}(\tau)=T_{\tau}\exp\left\{ -\frac{1}{\hbar}\int_{0}^{\tau}d\tau'\,{H}_{\rm C}(\tau')\right\},
\end{equation}
where $T_\tau$ is the imaginary-time time-ordering operator. For the free energy of the normal region this yields
\begin{equation}
F_{\rm N}=F_{0}-T\ln\langle{\cal S}\rangle_{0},
\end{equation}
where $\langle \dots \rangle_0$ implies taking a Gibbs statistical average over the unperturbed ground state (the subscript $0$ will be implied from now on).

All allowed diagrams, connected as well as disconnected, should be
taken into account when evaluating $\langle{\cal S}\rangle$. A combinatorial
exercise (see e.g.\ Chapter 15 in \cite{adg:books})
shows that the final result can be compactly expressed in terms of
connected diagrams only. With the fully connected
part of $\langle{\cal S}\rangle$ defined as
\begin{equation}
\langle{\cal S}\rangle_{{\rm con}}=1+\Xi_{1}+\Xi_{2}+\dots,
\end{equation}
with $\Xi_{n}$ representing all fully connected diagrams of order
$n$, it follows that
\begin{equation}
F_{\rm N}=F_{0}-T(\Xi_{1}+\Xi_{2}+\dots)=F_{0}-T(\langle{\cal S}\rangle_{{\rm con}}-1).
\end{equation}
The corrections thus directly follow from the series $\Xi_{n}$, up
to any desired particular order in the perturbation ${H}_{\rm C}$.

The leading order depending on the phase difference is second order
in the self-energy due to the proximity of the superconductors, i.e.,
fourth order in the coupling hamiltonian $H_{\rm C}$. We thus need
\begin{equation}
\Xi_{4}=\frac{1}{4!\hbar^4}\int_{0}^{\beta}\cdots\int_{0}^{\beta}d\tau_{1}\cdots d\tau_{4}\,\langle {T}_{\tau}H_{\rm C}(\tau_{1})H_{\rm C}(\tau_{2})H_{\rm C}(\tau_{3})H_{\rm C}(\tau_{4})\rangle.
\end{equation}
The only contributions
to $\Xi_{4}$ that know of the phases of both superconductors are
\begin{eqnarray}
\Xi_{4} & = & \frac{\gamma^{4}}{\hbar^4}\sum_{\sigma_{1..4}}\int \! dy_{1..4} \int_{0}^{\beta}\!\!\! d\tau_{1..4}\,\langle{T}_{\tau}{\psi}_{\sigma_{1}}^{\dagger}(x_L,y_{1};\tau_{1}){\Psi}_{\sigma_{1}}(x_L,y_{1};\tau_{1}){\psi}_{\sigma_{2}}^{\dagger}(x_L,y_{2};\tau_{2}){\Psi}_{\sigma_{2}}(x_L,y_{2};\tau_{2})\nonumber\\
{} & {} & \hspace{8em} \times {\Psi}_{\sigma_{3}}^{\dagger}(x_R,y_{3};\tau_{3}){\psi}_{\sigma_{3}}(x_R,y_{3};\tau_{3}){\Psi}_{\sigma_{4}}^{\dagger}(x_R,y_{4};\tau_{4}){\psi}_{\sigma_{4}}(x_R,y_{4};\tau_{4})\rangle\nonumber\\
 & {} & {} + {} \frac{\gamma^{4}}{\hbar^4}\sum_{\sigma_{1..4}}\int \! dy_{1..4} \int_{0}^{\beta}\!\!\! d\tau_{1..4}\,\langle{T}_{\tau}{\psi}_{\sigma_{1}}^{\dagger}(x_R,y_{1};\tau_{1}){\Psi}_{\sigma_{1}}(x_R,y_{1};\tau_{1}){\psi}_{\sigma_{2}}^{\dagger}(x_R,y_{2};\tau_{2}){\Psi}_{\sigma_{2}}(x_R,y_{2};\tau_{2})\nonumber\\
 & {} & {} \hspace{8em}\times {\Psi}_{\sigma_{3}}^{\dagger}(x_L,y_{3};\tau_{3}){\psi}_{\sigma_{3}}(x_L,y_{3};\tau_{3}){\Psi}_{\sigma_{4}}^{\dagger}(x_L,y_{4};\tau_{4}){\psi}_{\sigma_{4}}(x_L,y_{4};\tau_{4})\rangle.
\end{eqnarray}
From here on we leave out the subscript $L,R$ at the creation and annihilation operators for the electrons in the superconductors: The $x$-coordinate unambiguously implies which superconductor is involved. We now use Wick's theorem to write the eight-point correlation functions as products of two-point correlation functions. The result is expressed in terms of the Fourier transforms of the imaginary-time Green functions, defined by
\begin{eqnarray*}
-\langle{T}_{\tau}{\psi}_{\sigma_{1}}(x_{1},y_{1};\tau_{1}){\psi}^\dagger_{\sigma_{2}}(x_{2},y_{2};\tau_{2})\rangle & = & T\sum_{k}e^{-i\omega_{k}(\tau_{1}-\tau_{2})}{\cal G}(x_{1},y_{1},\sigma_{1};x_{2},y_{2},\sigma_{2};i\omega_{k}),\\
-\langle{T}_{\tau}{\Psi}_{\sigma_{1}}(x_{1},y_{1};\tau_{1}){\Psi}^\dagger_{\sigma_{2}}(x_{2},y_{2};\tau_{2})\rangle & = & T\sum_{k}e^{-i\omega_{k}(\tau_{1}-\tau_{2})}{\cal G}_{ee}^{{\rm sc}}(x_{1},y_{1},\sigma_{1};x_{2},y_{2},\sigma_{2};i\omega_{k}),\\
-\langle{T}_{\tau}{\Psi}_{\sigma_{1}}(x_{1},y_{1};\tau_{1}){\Psi}_{\sigma_{2}}(x_{2},y_{2};\tau_{2})\rangle & = & T\sum_{k}e^{-i\omega_{k}(\tau_{1}-\tau_{2})}{\cal G}_{eh}^{{\rm sc}}(x_{1},y_{1},\sigma_{1};x_{2},y_{2},\sigma_{2};i\omega_{k}),
\end{eqnarray*}
etc. Here, $\omega_k = \pi(2k+1)T/\hbar$ are the fermionic Matsubara frequencies. We thus find
\begin{eqnarray}
\Xi_{4} & = & \frac{\gamma^{4}T^{4}}{\hbar^4} \sum_{k_{1..4}}\sum_{\sigma_{1..4}}\int dy_{1..4} \int_{0}^{\beta}d\tau_{1..4}\\
 &  & \times\big[-e^{-i\omega_{k_{1}}(\tau_{1}-\tau_{2})}{\cal G}_{eh}^{{\rm sc}}(x_L,y_{1},\sigma_{1};x_L,y_{2},\sigma_{2};i\omega_{k_{1}})e^{-i\omega_{k_{2}}(\tau_{3}-\tau_{4})}{\cal G}_{he}^{{\rm sc}}(x_R,y_{3},\sigma_{3};x_R,y_{4},\sigma_{4};i\omega_{k_{2}}) \nonumber\\
 & & \hspace{2em}\times e^{-i\omega_{k_{3}}(\tau_{3}-\tau_{1})}{\cal G}(x_R,y_{3},\sigma_{3};x_L,y_{1},\sigma_{1};i\omega_{k_{3}})e^{-i\omega_{k_{4}}(\tau_{4}-\tau_{2})}{\cal G}(x_R,y_{4},\sigma_{4};x_L,y_{2},\sigma_{2};i\omega_{k_{4}})\nonumber\\
 &  & \hspace{1.5em}-e^{-i\omega_{k_{1}}(\tau_{1}-\tau_{2})}{\cal G}_{eh}^{{\rm sc}}(x_R,y_{1},\sigma_{1};x_R,y_{2},\sigma_{2};i\omega_{k_{1}})e^{-i\omega_{k_{2}}(\tau_{3}-\tau_{4})}{\cal G}_{he}^{{\rm sc}}(x_L,y_{3},\sigma_{3};x_L,y_{4},\sigma_{4};i\omega_{k_{2}})\nonumber\\
 & & \hspace{2em}\times e^{-i\omega_{k_{3}}(\tau_{3}-\tau_{1})}{\cal G}(x_L,y_{3},\sigma_{3};x_R,y_{1},\sigma_{1};i\omega_{k_{3}})e^{-i\omega_{k_{4}}(\tau_{4}-\tau_{2})}{\cal G}(x_L,y_{4},\sigma_{4};x_R,y_{2},\sigma_{2};i\omega_{k_{4}})\big].\nonumber
\end{eqnarray}

The anomalous Green functions in the superconductor follow from
\begin{align}
{\cal G}_{eh}^{{\rm sc}}(x,y_{1},\sigma_{1};x,y_{2},\sigma_{2};i\omega_{n}) = \delta_{\sigma_2,\bar\sigma_1} \frac{\sigma_1}{\cal A} \sum_{\bf k} e^{ik_y(y_1-y_2)} \frac{\Delta e^{i\varphi(x)}}{\omega_n^2+\varepsilon^2 + \Delta^2}
\end{align}
and similarly for ${\cal G}^{\rm sc}_{he}$. We write
\begin{align}
{\cal G}_{eh}^{{\rm sc}}(x,y_{1},\sigma_{1};x,y_{2},\sigma_{2};i\omega_{n}) = \delta_{\sigma_2,\bar\sigma_1} \frac{\sigma_1}{\cal A} \int d\varepsilon\, \sum_{\bf k} \delta(\varepsilon-\varepsilon_k) e^{ik_y(y_1-y_2)} \frac{\Delta e^{i\varphi(x)}}{\omega_n^2+\varepsilon_k^2 + \Delta^2}.
\end{align}
We assume that only $y_1 = y_2$ contributes significantly, and write in terms of the normal-state density of states
\begin{align*}
\nu_{\rm sc}(\varepsilon) = \frac{1}{\cal A} \sum_{\bf k} \delta(\varepsilon - \varepsilon_k),
\end{align*}
the expressions
\begin{eqnarray}
{\cal G}_{eh}^{{\rm sc}}(x,y_{1},\sigma_{1};x,y_{2},\sigma_{2};i\omega_{n}) & = & W\, \delta(y_{1}-y_{2}) \delta_{\sigma_{2},\bar{\sigma}_{1}}\sigma_{1}\frac{\pi\nu_{{\rm sc}}\Delta e^{i\varphi(x)}}{\sqrt{\Delta^{2}+(\hbar \omega_{n})^{2}}},\\
{\cal G}_{he}^{{\rm sc}}(x,y_{1},\sigma_{1};x,y_{2},\sigma_{2};i\omega_{n}) & = & -W\, \delta(y_{1}-y_{2})\delta_{\sigma_{2},\bar{\sigma}_{1}}\sigma_{1}\frac{\pi\nu_{{\rm sc}}\Delta e^{-i\varphi(x)}}{\sqrt{\Delta^{2}+(\hbar \omega_{n})^{2}}},
\end{eqnarray}
where $W$ is the width of the contact and $\nu_{{\rm sc}}$ is assumed to be constant for the energy range of interest. We further have $\varphi(x_L)=-\varphi/2$, and $\varphi(x_R)=\varphi/2$. Then we perform the integrals over all imaginary times, sum over all spins, and write the Josephson current as
\begin{eqnarray}
I^{(4)}_s(\varphi) & = & \frac{2e}{\hbar}( \gamma^{2}\pi\nu_{{\rm sc}}W)^{2}T \sum_{k} \sum_{\sigma_{1},\sigma_2} \int dy_1\,dy_2\, \frac{2\Delta^{2}}{\Delta^{2}+(\hbar\omega_{k})^{2}}\sigma_{1}\sigma_{2}\nonumber\\
& & \qquad\times {\rm Im}\big\{-e^{-i\varphi}{\cal G}_{\sigma_{2}\sigma_{1}}(x_R,y_{2};x_L,y_{1};i\omega_{k}){\cal G}_{\bar{\sigma}_{2}\bar{\sigma}_{1}}(x_R,y_{2};x_L,y_{1};-i\omega_{k})\big\},\label{eq:i4}
\end{eqnarray}
where the superscript $(4)$ indicates that this is the fourth-order contribution to the supercurrent (fourth order in $\gamma$). We see that this expression (ignoring spin) exactly reproduces Eq.\ (2.12) from Ref.~\cite{doi:10.1143/JPSJ.68.954s}.

The first relevant higher-order correction is eighth order in $\gamma$, and thus requires an evaluation of $\Xi_8$. The derivation is cumbersome but straightforward, and the result is
\begin{eqnarray}
I^{(8)}_s(\varphi) & = & \frac{2e}{\hbar}(\gamma^{2}\pi\nu_{{\rm sc}}W)^{4}T \sum_{k}  \sum_{\sigma_{1..4}} \int dy_{1..4}\,\frac{2\Delta^{4}}{[\Delta^{2}+(\hbar\omega_{k})^{2}]^{2}}\sigma_{1}\sigma_{2}\sigma_{3}\sigma_{4}\nonumber\\
& & \qquad\times {\rm Im} \big\{ 2e^{-2i\varphi}{\cal G}_{\sigma_{3}\sigma_{1}}(x_R,y_{3};x_L,y_{1};i\omega_{k}){\cal G}_{\sigma_{4}\bar{\sigma}_{1}}(x_R,y_{4};x_L,y_{1};-i\omega_{k})\nonumber\\
& & \hspace{7em} \times {\cal G}_{\bar{\sigma}_{3}\sigma_{2}}(x_R,y_{3};x_L,y_{2};-i\omega_{k}){\cal G}_{\bar{\sigma}_{4}\bar{\sigma}_{2}}(x_R,y_{4};x_L,y_{2};i\omega_{k})\nonumber\\
& & \hspace{5em} -e^{-i\varphi}{\cal G}_{\sigma_{3}\sigma_{1}}(x_R,y_{3};x_L,y_{1};i\omega_{k}){\cal G}_{\bar{\sigma}_{4}\bar{\sigma}_{1}}(x_R,y_{4};x_L,y_{1};-i\omega_{k})\nonumber\\
& & \hspace{7em} \times{\cal G}_{\bar{\sigma}_{3}\sigma_{2}}(x_R,y_{3};x_R,y_{2};-i\omega_{k}){\cal G}_{\sigma_{4}\bar{\sigma}_{2}}(x_R,y_{4};x_R,y_{2};i\omega_{k})\nonumber\\
& & \hspace{5em} -e^{-i\varphi}{\cal G}_{\sigma_{1}\sigma_{3}}(x_R,y_{1};x_L,y_{3};i\omega_{k}){\cal G}_{\bar{\sigma}_{1}\bar{\sigma}_{4}}(x_R,y_{1};x_L,y_{4};-i\omega_{k})\nonumber\\
& & \hspace{7em} \times{\cal G}_{\sigma_{2}\bar{\sigma}_{3}}(x_L,y_{2};x_L,y_{3};-i\omega_{k}){\cal G}_{\bar{\sigma}_{2}\sigma_{4}}(x_L,y_{2};x_L,y_{4};i\omega_{k})\big\}.\label{eq:i8}
\end{eqnarray}

\subsection{Numerical evaluation of the supercurrent}

\noindent The lattice Hamiltonian for the normal region is constructed as
\begin{eqnarray}
{H}_{\rm N} & = & \delta_{x_{1},x_{2}}\delta_{y_{1},y_{2}}\left[(4t-\mu_N+V_{x_1,y_1})\sigma_{0}+\tfrac{1}{2}g\mu_{{\rm B}}{\bf B}\cdot{\boldsymbol\sigma}\right]\nonumber\\
 &  & -\delta_{x_{1}+1,x_{2}}\delta_{y_{1},y_{2}}\left[e^{-2\pi iy_{1}\phi_{B}}\left(t\sigma_{0}+t_{xx}i\sigma_{x}+t_{xy}i\sigma_{y}\right)\right]\nonumber\\
 & & -\delta_{x_{1}-1,x_{2}}\delta_{y_{1},y_{2}}\left[e^{2\pi iy_{1}\phi_{B}}\left(t\sigma_{0}-t_{xx}i\sigma_{x}-t_{xy}i\sigma_{y}\right)\right]\nonumber\\
 &  & -\delta_{x_{1},x_{2}}\delta_{y_{1}+1,y_{2}}\left[t\sigma_{0}+t_{yx}i\sigma_{x}+t_{yy}i\sigma_{y}\right]\nonumber\\
 & & -\delta_{x_{1},x_{2}}\delta_{y_{1}-1,y_{2}}\left[t\sigma_{0}-t_{yx}i\sigma_{x}-t_{yy}i\sigma_{y}\right],
\end{eqnarray}
where $t=\hbar^{2}/2ma^{2}$ (with $a$ the lattice spacing), $\phi_{B}=B_za^{2}/\Phi_{0}$ (with $\Phi_0 = h/e$ the flux quantum), and the spin-orbit tunnel coupling elements are defined from
\begin{equation}
{H}_{{\rm SOI}}=\sum_{\alpha,\beta \in \{x,y\}}\frac{A_{\alpha\beta}}{\hbar} p_{\alpha}\sigma_{\beta},
\end{equation}
written now for convenience without the effect of a vector potential (in the numerical calculations the vector potential is included through a Peierls substitution, see $H_{\rm N}$ above).
With the ``regular'' coupling ${H}_{{\rm SOI}}=\frac{\alpha}{\hbar}(-p_{y}\sigma_{x}+p_{x}\sigma_{y})+\frac{\beta}{\hbar}(-p_{x}\sigma_{x}+p_{y}\sigma_{y})$
we have $A_{yy}=-A_{xx}=\beta$ and $A_{xy}=-A_{yx}=\alpha$. From
this coupling matrix we derive $t_{\alpha\beta}=A_{\alpha\beta}/2a$.

We then can use the expressions (\ref{eq:i4}) and (\ref{eq:i8}) to calculate the current. The integrals over the $y$-coordinates are replaced by sums and the factor $W$ drops from the prefactors. The remaining factor $\gamma^2 \pi \nu_{\rm sc}$ has dimensions of energy and corresponds to the (normal-state) tunneling rate of electrons into the superconductor at the Fermi energy. All required electronic Green functions then follow from solving for elements of
\begin{equation}
{\cal G}_{\sigma_{1},\sigma_{2}}(x_{1},y_{1};x_{2},y_{2};i\omega)=\left[\frac{1}{i\omega-{H}_{\rm N}}\right]_{x_{1},y_{1},\sigma_{1};x_{2},y_{2},\sigma_{2}}.
\end{equation}

With these numerically evaluated Green functions we then calculate the supercurrent using Eqs.~(\ref{eq:i4}) and (\ref{eq:i8}). For all plots in presented in the main text we assumed a temperature of $T = 100$~mK. We summed over Matsubara frequencies in a finite range, $k \in [-k_{\rm max},k_{\rm max}]$, where we chose $k_{\rm max}$ such that the sum converged within $\sim 10^{-3}$. For $T = 100$~mK and $\Delta = 0.1$~meV, we typically found $k_{\rm max} \approx 15$.


\begin{thebibliography}{99}
\bibitem{fraunhofer_theory:note1}
This form of the Hamiltonian implies that the $x$- and $y$-directions lie along
  the $[100]$ and $[010]$ crystallographic axes respectively. For a description
  with an arbitrary angle $\chi$ between the $x$-direction and the $[100]$-axis
  (still assuming $\hat z$ to point along the $[001]$-axis), one can use the
  modified Dresselhaus term $\frac{\beta}{\hbar}(-p_x\sigma_x + p_y\sigma_y)\cos 2\chi +
  \frac{\beta}{\hbar}(p_y\sigma_x + p_x\sigma_y)\sin 2\chi$, see~\cite{fraunhofer_theory:sup}.
\bibitem{fraunhofer_theory:note2}
In the absence of a perpendicular magnetic field, $B_z=0$, there are in fact 16
  symmetry transformations that fulfill $UH(\varphi)U^\dagger = H(-\varphi)$:
  See \cite{fraunhofer_theory:sup} for a complete overview.
\bibitem{footnote3}
The length was chosen to be larger than the spin-orbit length but shorter than the coherence length.
The width of the junction was practically limited to $\sim 35$ sites; for more sites the numerical calculations (especially the ones that are fourth order in $\kappa$) became too slow.
\end{thebibliography}
\end{document}